%% file: PROC-CTD2023-51.tex
\documentclass[10pt, paper=a4, UKenglish]{article}
\usepackage{graphicx}
\def\Title#1{\begin{center} {\Large #1 } \end{center}}
\def\Author#1{\begin{center}{ \sc #1} \end{center}}
\def\Address#1{\begin{center}{ \it #1} \end{center}}

\newcommand\pubblock{\rightline{\begin{tabular}{l} Proceedings of the CTD 2023\\ \pubnumber\\
         \pubdate  \end{tabular}}}

\newenvironment{Abstract}{\begin{quotation} \begin{center} 
             \large ABSTRACT \end{center}\bigskip 
      \begin{center}\begin{large}}{\end{large}\end{center} \end{quotation}}

\newenvironment{Presented}{\begin{quotation} \begin{center} 
             PRESENTED AT\end{center}\bigskip 
      \begin{center}\begin{large}}{\end{large}\end{center} \end{quotation}}

\def\Acknowledgements{\bigskip  \bigskip \begin{center} \begin{large}
      \bf ACKNOWLEDGEMENTS \end{large}\end{center}}

\input econfmacros.tex

\usepackage{color}
\usepackage{lineno}
\usepackage{subfig}
\usepackage[colorlinks,allcolors=blue]{hyperref}
\usepackage[capitalize]{cleveref}
\usepackage{floatrow}
\usepackage{mciteplus}
\mciteErrorOnUnknownfalse
\usepackage{ifthen}
\newboolean{articletitles}
\setboolean{articletitles}{true} 
\newboolean{uprightparticles}
\setboolean{uprightparticles}{false}
\input{lhcb-symbols-def}
\newcommand\pubnumber{PROC-CTD2023-51}

\newcommand\pubdate{November 30, 2023}

\def\affiliation{
  Aix Marseille Univ, CNRS/IN2P3, CPPM, Marseille, France}

\newcommand{\conference}{Connecting the Dots Workshop (CTD 2023)\\
October 10-13, 2023}

\usepackage{fancyhdr}
\definecolor{mygrey}{RGB}{105,105,105}
\begin{document}


\large
\begin{titlepage}
\pubblock

\vfill
\Title{First experiences with the LHCb heterogeneous software trigger}
\vfill

\Author{Andy Morris}
\Address{\affiliation}
\center{\textsc{On behalf of the} \,\lhcb \textsc{collaboration real time analysis group}}
\vfill

\begin{Abstract}
Since 2022, the \lhcb detector has been taking both proton-proton and lead-ion data at the LHC collision rate using a fully software-based trigger.
This has been implemented on GPUs at its first stage and CPUs at its second.
The setup allows for reconstruction, alignment, calibration and selections to be performed online -- known as the real time analysis paradigm.
As well as this, physics analyses are performed using the output of online reconstruction with early results shown using data taken in 2022.
\end{Abstract}

\vfill

\begin{Presented}
\conference
\end{Presented}
\vfill
\end{titlepage}
\def\thefootnote{\fnsymbol{footnote}}
\setcounter{footnote}{0}
%

\normalsize 


\input{text/introduction}
\input{text/detector}
\input{text/overview}
\input{text/alignment}
\input{text/HLT1}
\input{text/HLT2}
\input{text/conclusions}

\Acknowledgements
The author acknowledges the support of the European Research Council Starting grant ALPACA 101040710.


\input{references.bbl}


\end{document}

%% file: econfmacros.tex
\def\beq{\begin{equation}}
\def\eeq#1{\label{#1}\end{equation}}
\def\eeqn{\end{equation}}

\def\beqa{\begin{eqnarray}}
\def\eeqa#1{\label{#1}\end{eqnarray}}
\def\eeqan{\end{eqnarray}}

\let\bar=\overbar

\def\eg{{\it e.g.}}

\def\D{{\cal D}}

\def\Dslash{\not{\hbox{\kern-4pt $D$}}}
\def\dslash{\not{\hbox{\kern-2pt $\del$}}}

\def\msb{{\bar{\ssstyle M \kern -1pt S}}}



%% file: lhcb-symbols-def.tex
\usepackage{xspace} 
\usepackage{upgreek}

\def\lhcb   {\mbox{LHCb}\xspace}

\def\lhc    {\mbox{LHC}\xspace}

\def\velo   {VELO\xspace}
\def\rich   {RICH\xspace}
\def\richone {RICH1\xspace}
\def\richtwo {RICH2\xspace}

\def\ut       {UT\xspace}
\def\scifi    {SciFi\xspace}

\def\ecal   {ECAL\xspace}
\def\hcal   {HCAL\xspace}
\def\MagUp {\mbox{\em Mag\kern -0.05em Up}\xspace}

\def\lone   {L0\xspace}

\def\hltone {HLT1\xspace}
\def\hlttwo {HLT2\xspace}

\ifthenelse{\boolean{uprightparticles}}%
{

 \def\Pmu         {\ensuremath{\upmu}\xspace}

 \def\Ppi         {\ensuremath{\uppi}\xspace}

 \def\Ppsi        {\ensuremath{\uppsi}\xspace}

 \def\PDelta      {\ensuremath{\Delta}\xspace}                 
 \def\PXi         {\ensuremath{\Xi}\xspace}                 
 \def\PLambda     {\ensuremath{\Lambda}\xspace}                 
 \def\PSigma      {\ensuremath{\Sigma}\xspace}                 
 \def\POmega      {\ensuremath{\Omega}\xspace}                 
 \def\PUpsilon    {\ensuremath{\Upsilon}\xspace}
 \let\oldPi\Pi
 \def\PPi         {\ensuremath{\oldPi}\xspace}

 \def\PB      {\ensuremath{\mathrm{B}}\xspace}                 
                  
 \def\PD      {\ensuremath{\mathrm{D}}\xspace}

 \def\PJ      {\ensuremath{\mathrm{J}}\xspace}                 
 \def\PK      {\ensuremath{\mathrm{K}}\xspace}

 \def\Pb      {\ensuremath{\mathrm{b}}\xspace}

 \def\Pe      {\ensuremath{\mathrm{e}}\xspace}

 \def\Pi      {\ensuremath{\mathrm{i}}\xspace}

 \def\Pp      {\ensuremath{\mathrm{p}}\xspace}

 \def\Ps      {\ensuremath{\mathrm{s}}\xspace}

 \def\thebaroffset{0.0em}
}
{

 \def\Pmu         {\ensuremath{\mu}\xspace}

 \def\Ppi         {\ensuremath{\pi}\xspace}

 \def\Ppsi        {\ensuremath{\psi}\xspace}                 
                  
 \mathchardef\PDelta="7101
 \mathchardef\PXi="7104
 \mathchardef\PLambda="7103
 \mathchardef\PSigma="7106
 \mathchardef\POmega="710A
 \mathchardef\PUpsilon="7107
 \mathchardef\PPi="7105
                  
 \def\PB      {\ensuremath{B}\xspace}                 
                  
 \def\PD      {\ensuremath{D}\xspace}

 \def\PJ      {\ensuremath{J}\xspace}                 
 \def\PK      {\ensuremath{K}\xspace}

 \def\Pb      {\ensuremath{b}\xspace}

 \def\Pe      {\ensuremath{e}\xspace}

 \def\Pi      {\ensuremath{i}\xspace}

 \def\Pp      {\ensuremath{p}\xspace}

 \def\Ps      {\ensuremath{s}\xspace}

 \def\thebaroffset{0.18em}
}
\newcommand{\offsetoverline}[2][\thebaroffset]{\kern #1\overline{\kern -#1 #2}}%

\makeatletter
\ifcase \@ptsize \relax
  \newcommand{\miniscule}{\@setfontsize\miniscule{4}{5}}
\or
  \newcommand{\miniscule}{\@setfontsize\miniscule{5}{6}}
\or
  \newcommand{\miniscule}{\@setfontsize\miniscule{5}{6}}
\fi
\makeatother

\DeclareRobustCommand{\optbar}[1]{\shortstack{{\miniscule (\rule[.5ex]{1.25em}{.18mm})}
  \\ [-.7ex] $#1$}}

\def\en         {{\ensuremath{\Pe^-}}\xspace}   
\def\ep         {{\ensuremath{\Pe^+}}\xspace}

\def\mup        {{\ensuremath{\Pmu^+}}\xspace}
\def\mun        {{\ensuremath{\Pmu^-}}\xspace} 

\def\squark    {{\ensuremath{\Ps}}\xspace}

\def\bquark    {{\ensuremath{\Pb}}\xspace}

\def\pion   {{\ensuremath{\Ppi}}\xspace}

\def\pip    {{\ensuremath{\pion^+}}\xspace}
\def\pim    {{\ensuremath{\pion^-}}\xspace}
\def\pipm   {{\ensuremath{\pion^\pm}}\xspace}

\def\kaon    {{\ensuremath{\PK}}\xspace}

\def\KorKbar {\kern \thebaroffset\optbar{\kern -\thebaroffset \PK}{}\xspace}

\def\Kp      {{\ensuremath{\kaon^+}}\xspace}
\def\Km      {{\ensuremath{\kaon^-}}\xspace}

\def\Kmp     {{\ensuremath{\kaon^\mp}}\xspace}
\def\KS      {{\ensuremath{\kaon^0_{\mathrm{S}}}}\xspace}

\def\KL      {{\ensuremath{\kaon^0_{\mathrm{L}}}}\xspace}

\def\Dbar    {{\ensuremath{\offsetoverline{\PD}}}\xspace}
\def\D       {{\ensuremath{\PD}}\xspace}

\def\DorDbar {\kern \thebaroffset\optbar{\kern -\thebaroffset \PD}\xspace}
\def\Dz      {{\ensuremath{\D^0}}\xspace}
\def\Dzb     {{\ensuremath{\Dbar{}^0}}\xspace}
\def\Dp      {{\ensuremath{\D^+}}\xspace}
\def\Dm      {{\ensuremath{\D^-}}\xspace}

\def\DpDm    {\ensuremath{\Dp {\kern -0.16em \Dm}}\xspace}

\def\Dstarp  {{\ensuremath{\D^{*+}}}\xspace}

\def\B       {{\ensuremath{\PB}}\xspace}

\def\BorBbar {\kern \thebaroffset\optbar{\kern -\thebaroffset \PB}\xspace}

\def\Bd      {{\ensuremath{\B^0}}\xspace}

\def\BdorBdbar {\kern \thebaroffset\optbar{\kern -\thebaroffset \Bd}\xspace}
\def\Bu      {{\ensuremath{\B^+}}\xspace}

\def\Bp      {{\ensuremath{\Bu}}\xspace}

\def\Bs      {{\ensuremath{\B^0_\squark}}\xspace}

\def\BsorBsbar {\kern \thebaroffset\optbar{\kern -\thebaroffset \Bs}\xspace}

\def\jpsi     {{\ensuremath{{\PJ\mskip -3mu/\mskip -2mu\Ppsi}}}\xspace}

\def\Y#1S{\ensuremath{\PUpsilon{(#1S)}}\xspace}

\def\proton      {{\ensuremath{\Pp}}\xspace}
\def\pp          {\proton-\proton}

\def\Lz          {{\ensuremath{\PLambda}}\xspace}

\def\LorLbar     {\kern \thebaroffset\optbar{\kern -\thebaroffset \PLambda}\xspace}

\def\to                 {\ensuremath{\rightarrow}\xspace}

\def\AT#1     {\ensuremath{A_{\mathrm{T}}^{#1}}\xspace}           

\def\C#1      {\ensuremath{\mathcal{C}_{#1}}\xspace}                       
\def\Cp#1     {\ensuremath{\mathcal{C}_{#1}^{'}}\xspace}                    
\def\Ceff#1   {\ensuremath{\mathcal{C}_{#1}^{\mathrm{(eff)}}}\xspace}        
\def\Cpeff#1  {\ensuremath{\mathcal{C}_{#1}^{'\mathrm{(eff)}}}\xspace}       
\def\Ope#1    {\ensuremath{\mathcal{O}_{#1}}\xspace}                       
\def\Opep#1   {\ensuremath{\mathcal{O}_{#1}^{'}}\xspace}                    

\newcommand{\aunit}[1]{\ensuremath{\text{\,#1}}}       

\newcommand{\tev}{\aunit{Te\kern -0.1em V}\xspace}
\newcommand{\gev}{\aunit{Ge\kern -0.1em V}\xspace}
\newcommand{\mev}{\aunit{Me\kern -0.1em V}\xspace}
\newcommand{\kev}{\aunit{ke\kern -0.1em V}\xspace}
\newcommand{\ev}{\aunit{e\kern -0.1em V}\xspace}
 
\newcommand{\mevc}{\ensuremath{\aunit{Me\kern -0.1em V\!/}c}\xspace}
\newcommand{\gevc}{\ensuremath{\aunit{Ge\kern -0.1em V\!/}c}\xspace}
\newcommand{\mevcc}{\ensuremath{\aunit{Me\kern -0.1em V\!/}c^2}\xspace}
\newcommand{\gevcc}{\ensuremath{\aunit{Ge\kern -0.1em V\!/}c^2}\xspace}

\def\gsim{{~\raise.15em\hbox{$>$}\kern-.85em
          \lower.35em\hbox{$\sim$}~}\xspace}
\def\lsim{{~\raise.15em\hbox{$<$}\kern-.85em
          \lower.35em\hbox{$\sim$}~}\xspace}

\def\tell1  {TELL1\xspace}
\def\ukl1   {UKL1\xspace}

\def\cfourften     {\ensuremath{\mathrm{ C_4 F_{10}}}\xspace}
\def\cffour        {\ensuremath{\mathrm{ CF_4}}\xspace}

\newcommand{\lhcborcid}[1]{\href{https://orcid.org/#1}{\hspace*{0.1em}\raisebox{-0.45ex}{\includegraphics[width=1em]{figs/orcidIcon.pdf}}}}


%% file: text/introduction.tex
\section{Introduction}

Starting in 2022, the \lhcb experiment at the Large Hadron Collider (\lhc) has begun its first data taking periods (Run~3) following its upgrade~\cite{alves2008lhcb,LHCb:2023hlw}.
A key part of this upgrade is that \lhcb will take data from proton-proton (\pp) collisions with an average pile-up of $\mu = 5$, rather than its pre-upgrade average of $\mu = 1$, and as such a new trigger system has been developed to process the increased output data rate -- $5\,\mathrm{TB/s}$ -- due to the increase in instantaneous luminosity.
The trigger has also been developed to use the upgraded hardware; almost all subdetectors having been improved from their Run~2 equivalents, most of them being overhauled entirely.
The trigger is a two-stage process, with stages called `high-level-trigger 1(2)' -- \hltone (\hlttwo). The first of these two stages runs heterogeneously on Nvidia RTX A5000 GPUs, whereas the second stage runs on CPUs.
Notably, the \lhcb trigger has no hardware stage whatsoever, mitigating the inefficiency of the \lone trigger used during Runs~1 and 2.
As well as this, the role of alignment and calibration of the detector has been paramount during the commissioning process at the start of Run~3 as well as in data taking beyond.
The first experiences and results of the new trigger system as well as the alignment and calibration of the detector are outlined in these proceedings.

%% file: text/detector.tex
\section{Detector}

\begin{figure}[b]
    \centering
    \includegraphics[width=0.54\linewidth]{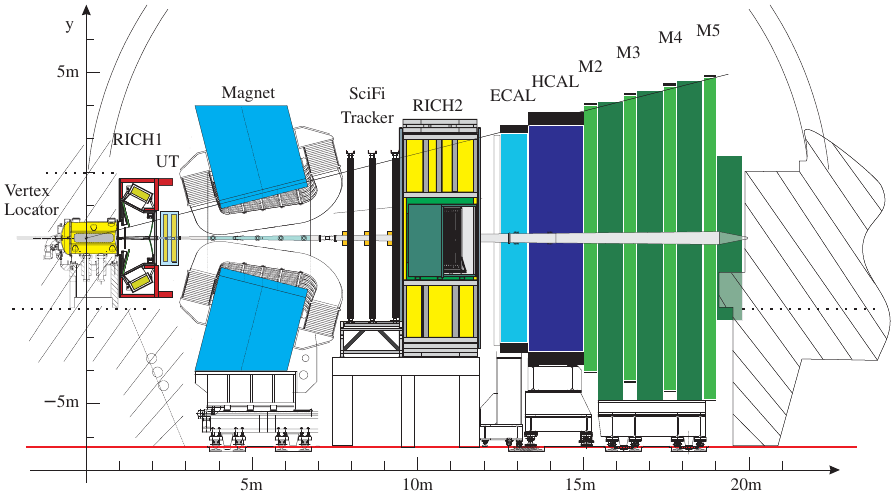}
    \caption{A schematic diagram of the upgraded \lhcb detector.}
    \label{fig:LHCb}
\end{figure}

\lhcb is a single-arm forward spectrometer covering the pseudorapidity range
$2 < \eta < 5$, located at interaction point number 8 on the \lhc ring.
\Cref{fig:LHCb} shows the layout of the upgraded detector.
The coordinate system used throughout this paper has the origin at the nominal \pp interaction point, the $z$ axis along the beam pointing downstream, the $y$ axis pointing vertically upward and the $x$ axis towards the \lhc center.

The particle tracking system comprises an array of pixel silicon detectors 
surrounding the interaction region called the \velo, the silicon-strip \ut
in front of the large-aperture dipole magnet, and three \scifi stations
downstream of the magnet.
The \ut completely replaces the Tracker Turencis used in Runs~1 and 2 with the \scifi replacing the T-trackers.
It should be noted that during 2022, the \ut had not completed its installation and during 2023 it was not used in data taking.
Furthermore, during 2023, the \velo subdetector was operated with a $49\,\mathrm{mm}$ opening due to a deformation in its RF foil ~\cite{StatusReport}.
The RF foil will be replaced during the 2023 year end technical stop, and the \ut is expected to be full operational in 2024.

Particle identification (PID) for charged particles is provided by two ring imaging Cherenkov detectors (\richone and \richtwo) using \cfourften and \cffour gases as radiators.
Compared to those used in Runs~1 and 2, the \richone subdetector has entirely new optical systems designed to be optimised to the higher occupancies, with both \rich subdetectors having improved readout infrastructure to account for the higher luminosity.
As well as this, there are two calorimeter systems: the electromagnetic calorimeter (\ecal), providing PID for electrons and photons; and the hadronic calorimeter (\hcal), providing PID for all hadrons, especially \KL and neutrons. These have largely remained unchanged in the upgrade except for their front-end and readout electronics.
Finally, there are four muon stations (M2-M5), interleaved with iron shielding, to identify muons which can penetrate beyond the calorimeters, again with redesigned monitoring and control electronics.

Between the \ut and \scifi trackers is a large dipole magnet providing a magnetic field with a bending power of $4$\,Tm\xspace for charged particles passing through it. The polarity of this magnet is reversed (MagUp and MagDown) every few weeks.

The data acquisition (DAQ) system from the front end and back end electronics feeds to an event filter farm through radiation-resistant long-distance optical links. It is in this farm that the trigger is run.

%% file: text/overview.tex
\section{Data flow overview}

The development of the trigger system as well as the alignment and calibration are the main goals of the Real Time Analysis project at \lhcb, who manage the processing of \lhcb data in Run~3 and beyond.

The flow of data taking in \lhcb begins in the \hltone trigger.
Here, data is reconstructed at $30\,\mathrm{MHz}$ and filtered using the selection criteria of $\mathcal{O}(10)$ \textit{inclusive} trigger lines into general categories reducing the data rate by a factor of 30 and is outputted to a $\sim 30\,\mathrm{PB}$ buffer.
At this stage, the full detector output (the `raw banks') is persisted along with information about which lines passed by each persisted event, and which reconstructed candidates were responsible for each passing line.

The contents of this buffer is then used in two different ways.
Firstly, the buffer is used for performing the detector alignment.
This refers to corrections in the true size, shape, position and rotation of each piece of each subdetector compared to the values provided by the technical drawings.
Accurately finding these values allows for improvements in all aspects of the detector, notably tracking and PID.
Alignment is performed in real time, feeding back to \hltone so that as data is taken, the alignment improves which then improves the quality of future data, upon which the alignment may be further improved. Once alignment and calibration is of sufficient quality, it is then also used in the running of \hlttwo.

\begin{figure}[p]
    \centering
    \includegraphics[width=0.65\linewidth]{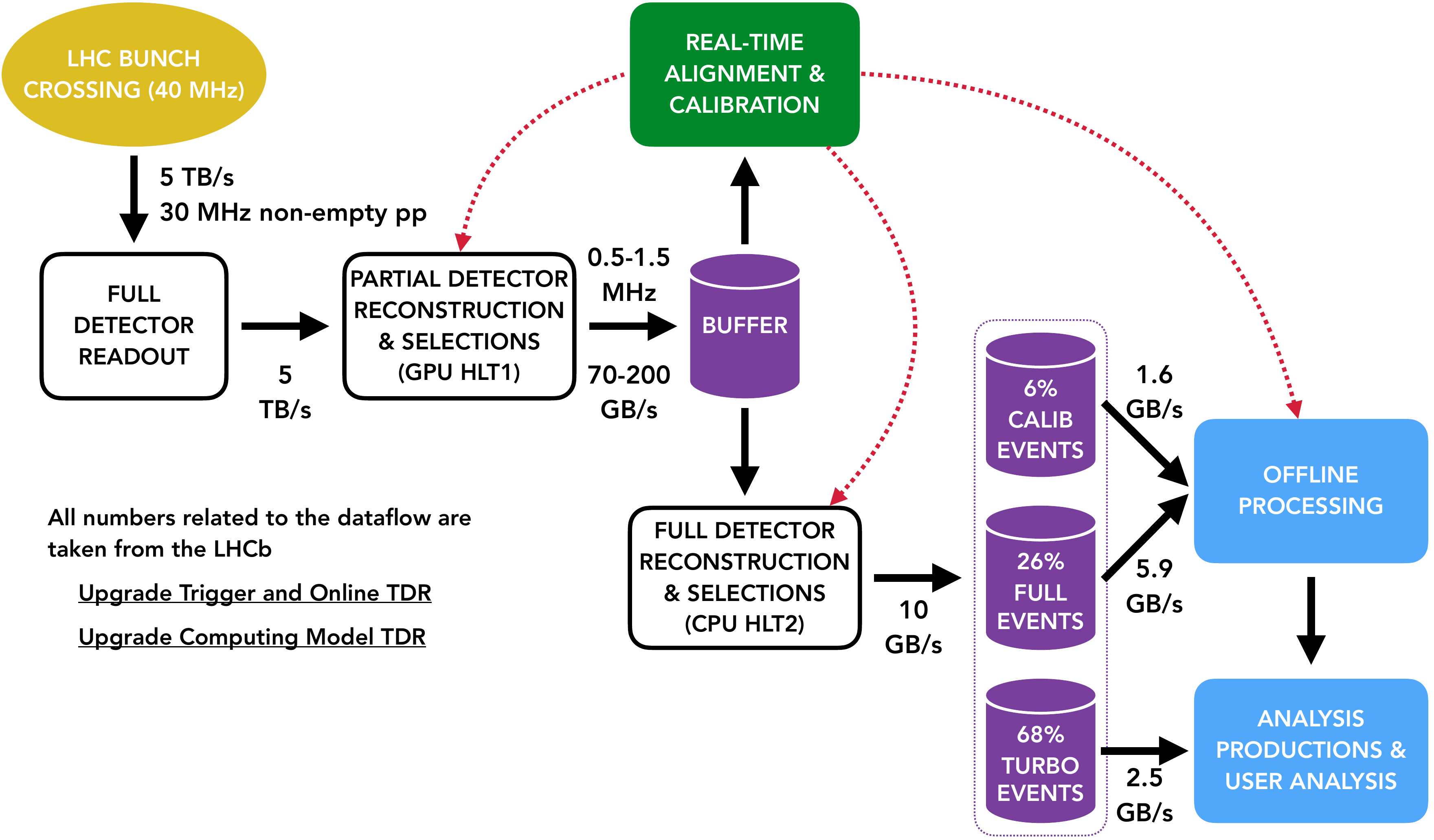}
    \caption{The data flow of \lhcb from raw detector output through the trigger \cite{LHCb-FIGURE-2020-016}.}
    \label{fig:dataflow}
\end{figure}

The second way in which the output of the buffer is used is as the input for \hlttwo.
This trigger, in contrast to \hltone, processes the data into $\mathcal{O}(1000)$ \textit{exclusive} lines, each of which is typically specific to a single (or few) decay channels to be used by analysists.
These selections use the full offline-quality reconstruction as well as all available PID information so that as much selection as possible may be performed at this stage (and not offline), reducing bandwidth.
At this stage, the raw banks typically are not persisted (though they can be upon request), rather only `physics objects' are persisted -- the reconstructed tracks considered by the trigger to be the `signal'.
In contrast to \hlttwo in Run~1 and Run~2 at \lhcb, it is expected that analysists perform as much of their selection as possible directly in their line, doing only minimal selection offline, minimising the amount of storage space to be used unnecessarily.
The final output of \hlttwo to storage, due to these efforts, is reduced to only 10\,GB/s. A flowchart summarising the data flow is given in \cref{fig:dataflow}.

%% file: text/alignment.tex
\section{Alignment and calibration performance}

Over the first year of \lhcb data taking since its upgrade, vast gains have been made in track reconstruction efficiency and PID response due to the improvement of alignment as well as to the subdetectors directly.

\begin{figure}[p]
    \begin{floatrow}
        \ffigbox{
            \includegraphics[width=\linewidth]{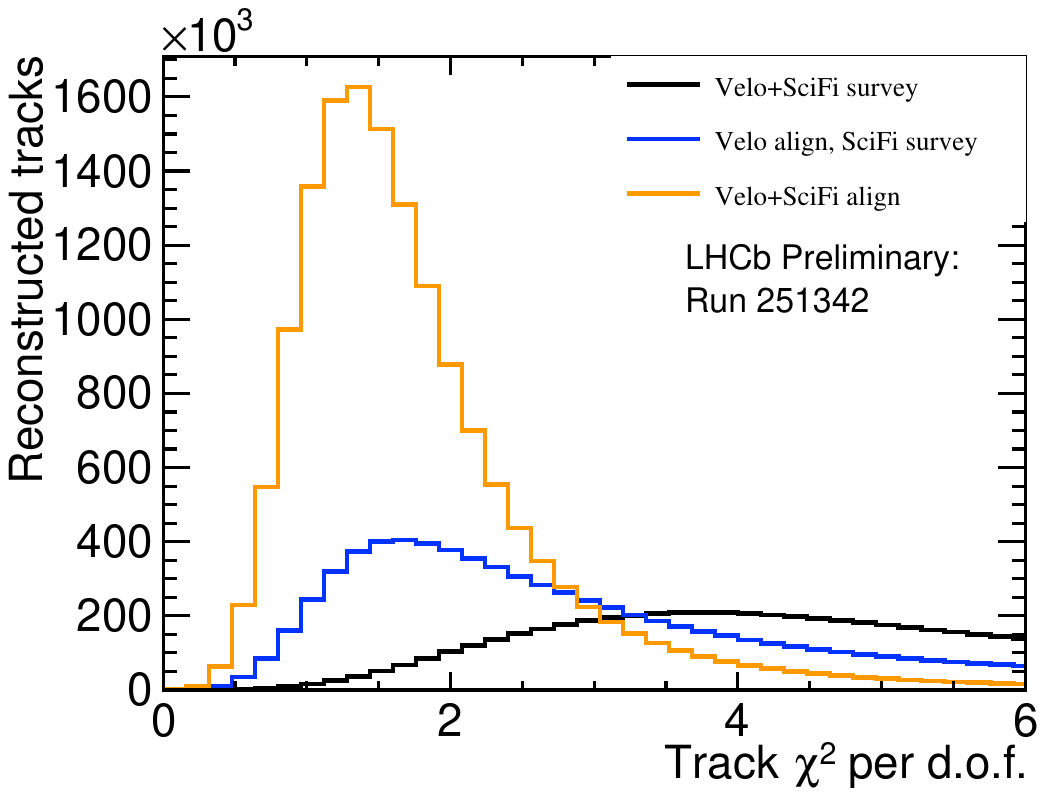}
        }{
            \caption{The quality of fitted long tracks before and after the alignment of both the \velo and \scifi trackers~\cite{LHCb-FIGURE-2022-018}.}
            \label{fig:alignment1}
        }
        \ffigbox{
            \includegraphics[width=0.95\linewidth]{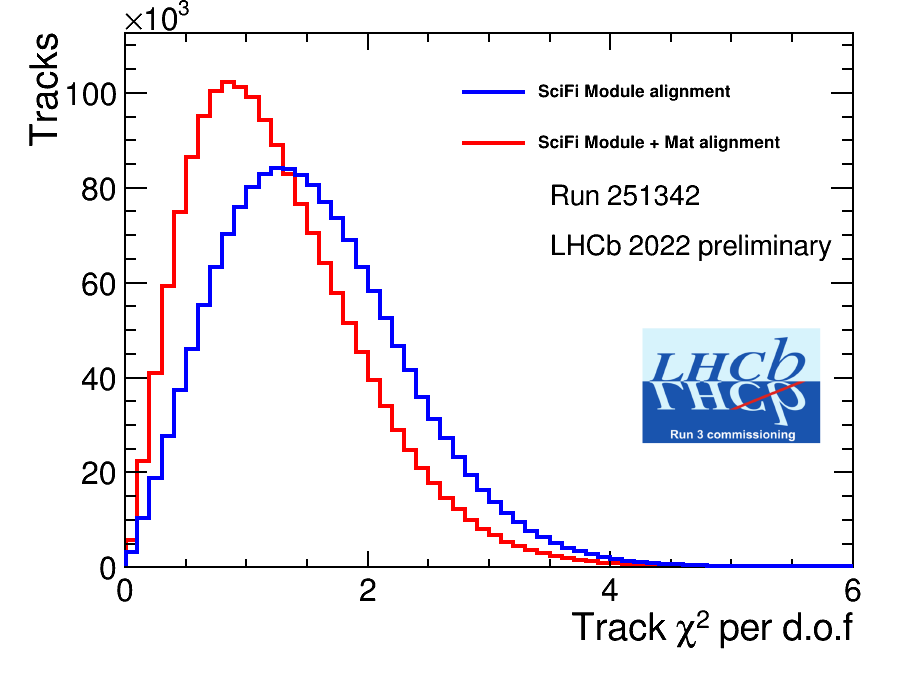}
        }{
            \caption{The further improvement in fitted track quality after the alignment of mats within the \scifi~\cite{LHCb-FIGURE-2023-018}.}
            \label{fig:alignment2}
        }
    \end{floatrow}
\end{figure}

The improvement in quality is given quantitatively by investigating the distribution of $\chi^2$ per degree of freedom for fits made by the Kalman filter in \hlttwo to long tracks -- that is tracks which span the full length of \lhcb.
For tracking, this has been done in two steps during the commissioning of process, firstly aligning the \velo and \scifi detector elements, shown in \cref{fig:alignment1}.
Each \scifi module then contains four detection planes (mats); aligning the mats within each \scifi element after aligning the elements as a whole produces greater tracking quality still, shown in \cref{fig:alignment2}.

\begin{figure}[p]
        \includegraphics[width=0.49\linewidth]{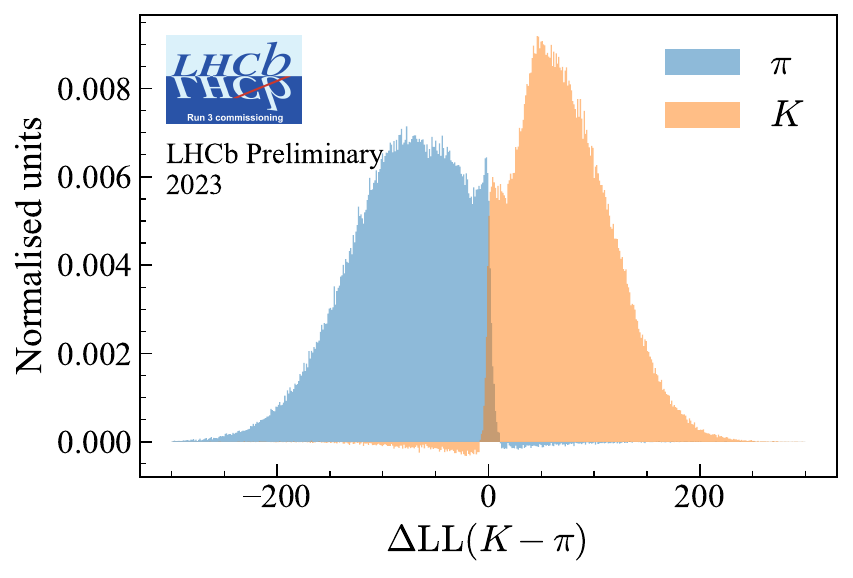}
        \includegraphics[width=0.49\linewidth]{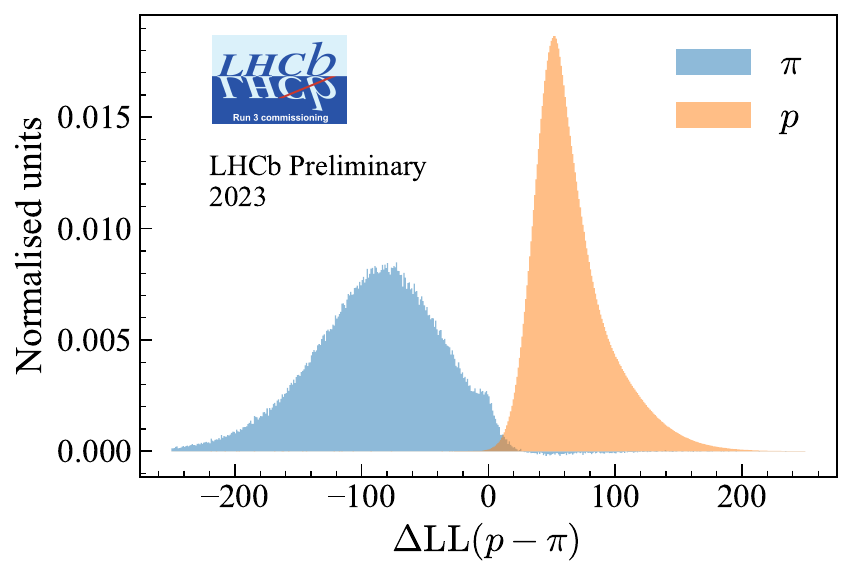}
        \caption{The difference between the log likelihood distributions for kaons and pions (left) as well as for protons and pions (right)~\cite{LHCb-FIGURE-2023-023}.}
        \label{fig:alignmentPID}
\end{figure}

Improvements to the PID response for hadrons with respect to Run~2 come from the excellent performance of the upgraded \rich detectors, such as better single-photon resolution and yield.
These are tested using dedicated calibration data samples.
For pions and kaons, $\Dstarp\to(\Dzb\to\Kp\pim)\pip$ decays are used, whereas for protons $\Lz\to\proton\pim$ are used: isolated from background by use of \textit{s\kern-0.1em{Weights}}~\cite{Pivk:2004ty}.
The separation of kaons and protons with respect to pions may be seen in \cref{fig:alignmentPID} where the log likelihoods of each candidate compared to a pion is given.
The separation between particle flavours is better than in Run~2 for comparable occupancies.
Further improvement is required for the higher occupancies of nominal Run~3 data taking conditions.

%% file: text/HLT1.tex
\section{\hltone details and results}

The first level trigger being able to reconstruct events at the full detector output rate is an achievement which should not be understated, however to do this it must make some concessions.
The main limiting factor of the first trigger is maintaining a high throughput, since it must process the data in real time. For this reason, the level of reconstruction it can perform is quite simplified compared to that of \hlttwo.
The first example of this is in the simplified track fitting compared to \hlttwo.
To maintain high-throughput, the Kalman filtering process on track uses a parameterised approach compared to the more robust approach of \hlttwo.
Furthermore, only the segment of a track inside the \velo is considered for the filtering process.
As well as this a more simplifed pattern recognition algorithm is used.

To find long tracks two algorithms are used called `forward tracking' and `seeding-matching'.
Forward tracking refers to taking \velo tracks and then tracing them forwards to the next tracker.
This is done linearly between the \velo and the magnet, we then approximate the magnetic field as a single shift in angle, in the middle of the magnet region to extrapolate to the \scifi.\footnote{We neglect the \ut from this description as its commissioning is still ongoing.}
This shift in the angle is calculated from the position of the tracks, their kinematic properties and the parameterisation of the magnetic field.
During each of these extrapolations, hits are found in the corresponding tracker and if enough hits are found for a track with a sufficiently low $\chi^2$ then it is kept~\cite{Amhis:1641927}.
On the other hand the seeding-matching algorithm instead creates `seed' tracks using hits only in the \scifi tracker.
\velo segment tracks are then extrapolated forwards towards the magnet region.
The \scifi seeds are then extrapolated back to the center of the magnet region and if they match the \velo segments, they are considered to form a single long track~\cite{Jashal:2826068}.
The current approach of \hltone is first find tracks using the forward algorithm, and then use a second pass to find the remaining tracks with the seeding-matching algorithm only using the remaining hits in each tracker after those used for forward tracking have been removed.

The second example in which the reconstruction is simplified is in its PID. The main PID inputs for hadrons in \lhcb come from the \rich subdetectors; these are ignored entirely at this stage.
Rather, the only PID information comes from the muon stations (to identify muons) and the \ecal (to identify electrons and photons), with all other particles assumed to be pions.

\begin{figure}[h]
    \centering
    \includegraphics[width=0.60\linewidth]{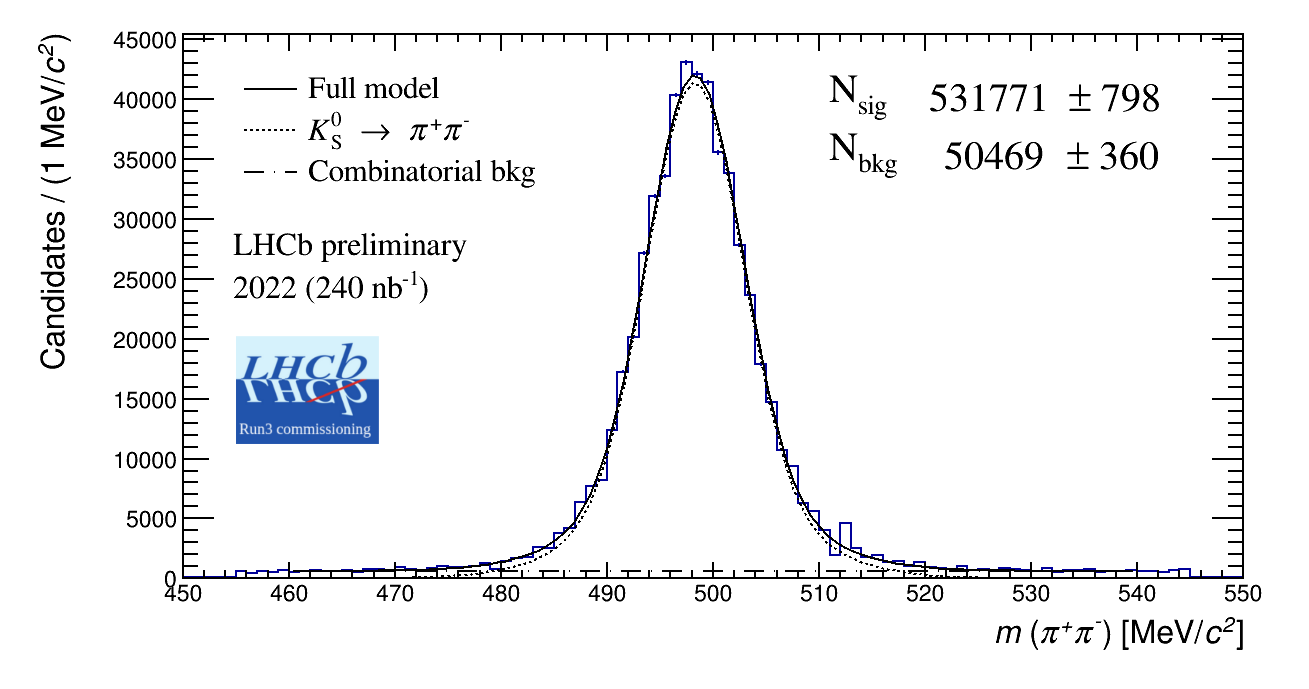}
    \includegraphics[width=0.39\linewidth]{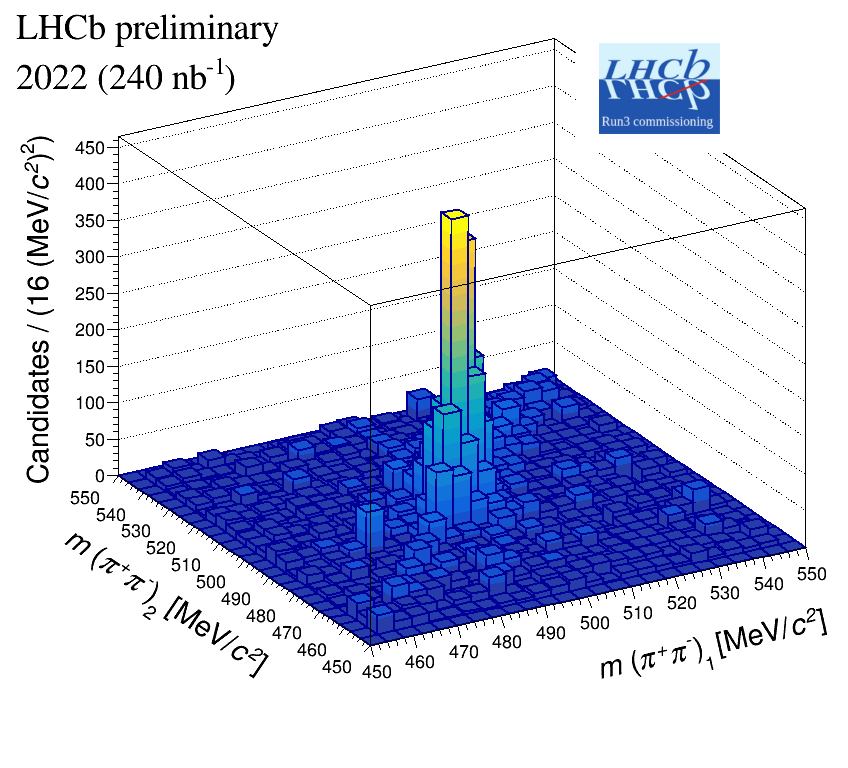}
    \caption{Mass plots of \KS candidates identified by \hltone's \texttt{TwoTrackKs} line for all candidates concatenated (left) and both candidates for each event separately (right)~\cite{LHCb-FIGURE-2023-005}.}
    \label{fig:KS}
\end{figure}

Despite these simplifications required to maintain high throughput, the resulting data taken by \hltone is already good enough to find clear mass peaks which may be used for physics.
The first example of this is taken from the inclusive line \texttt{TwoTrackKs}, which isolates events which are believed to contain two $\KS \to \pip\pim$ decays -- intended to be used \eg~for analyses of $\Dz \to \KS\KS$ decays.
The resulting mass peaks which can be found using only information from \hltone for both a single \KS candidate as well as the pair may be seen in \cref{fig:KS}, with a resolution approaching that of Run~2 in the analysis of $\Dz\to\KS\KS$~\cite{LHCb:2021rdn}.

\begin{figure}[t]
    \centering
    \includegraphics[width=0.49\linewidth]{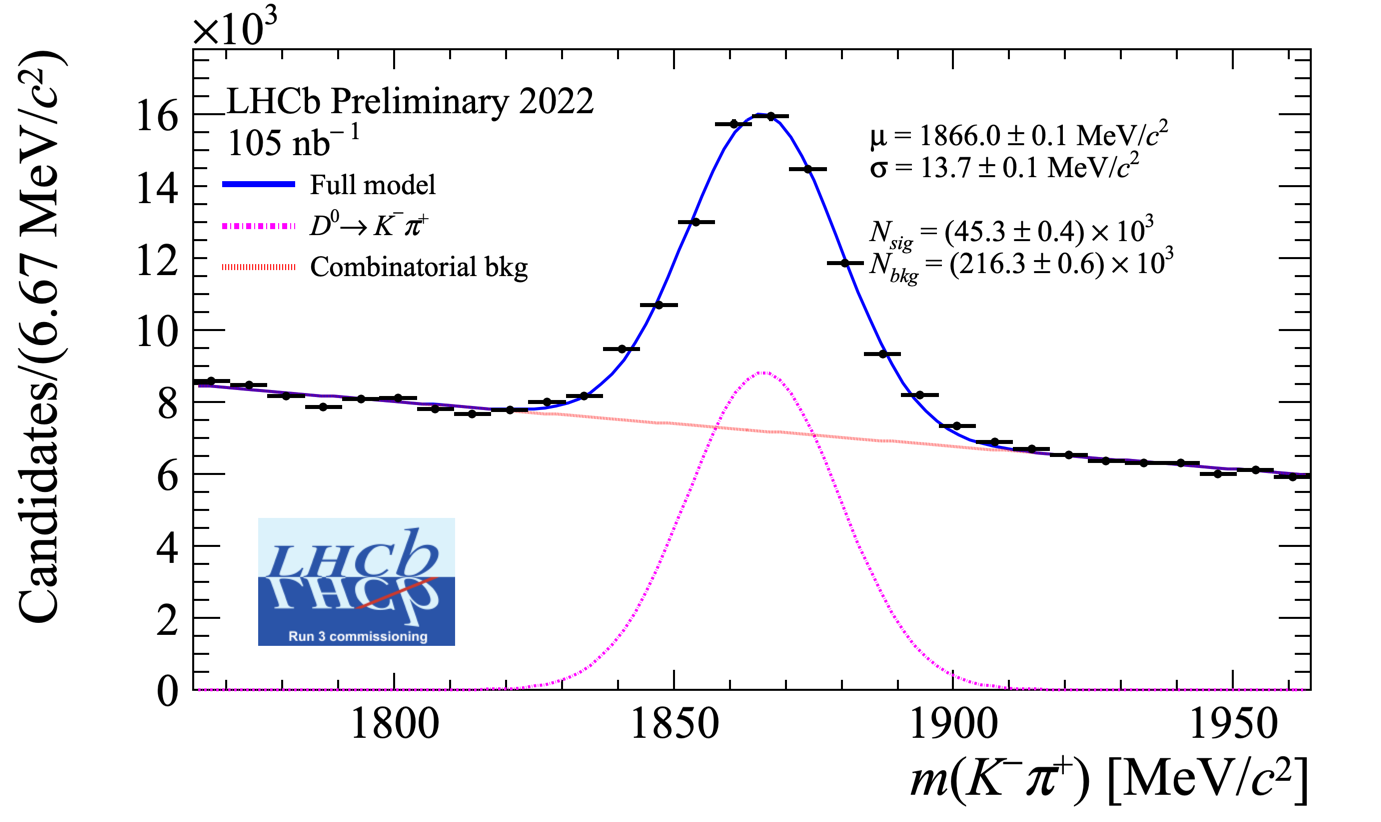}
    \includegraphics[width=0.49\linewidth]{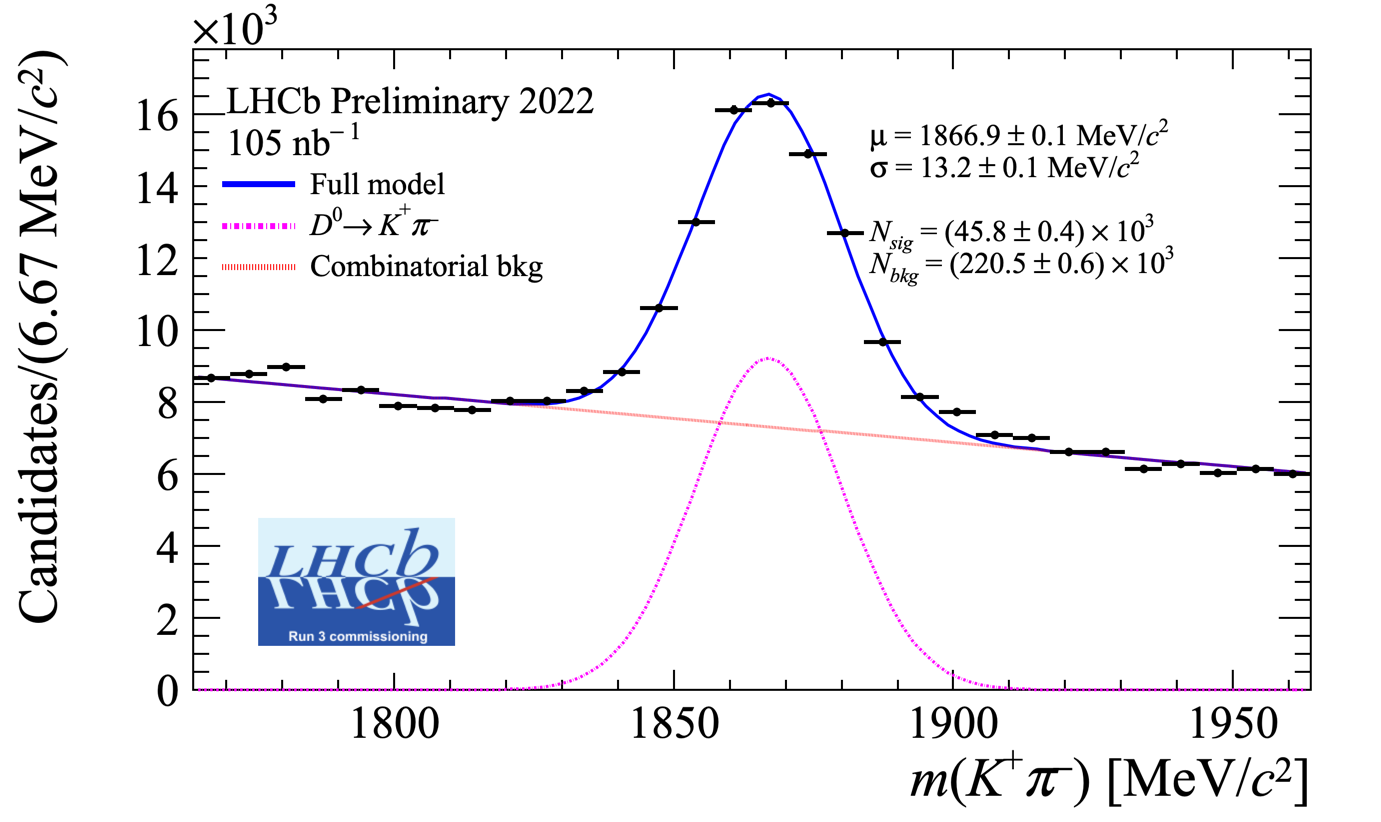}
    \caption{The resulting mass peaks for \Dz (left) and \Dzb (right) candidates found in \hltone~\cite{LHCb-FIGURE-2023-009}.}
    \label{fig:D0}
\end{figure}

As well as reconstructing pions into heavier parent particles, it is also possible to use other hadrons despite the lack of PID.
An example of this is in the decay $\DorDbar\,\kern-0.25em{{}^0} \to \Kmp\pipm$.
Here one of the pions (selected arbitrarily) may be assigned the mass of a kaon to find the \Dz peak.
The resulting mass peaks may be seen for both \Dz and \Dzb decays in \cref{fig:D0} where importantly, similar numbers of each particle flavour may be found in the same dataset; implying a lack of bias in the trigger selection.

A key part of the trigger software, \texttt{Allen}~\cite{Allen}, as well as performing the filtering of input events, is in monitoring the data being taken.
Monitoring allows for data quality to be ensured in real time and at the full $30\,\mathrm{MHz}$ decay rate, flagging up issues immediately to experts on-call 24-7 if deviations in the monitored variables start to appear.
Since both $\KS \to \pip\pim$ and $\DorDbar\,\kern-0.25em{{}^0}\to\Kmp\pipm$ mass peaks are visible directly to \hltone, these mass peaks are included as monitoring variables as well as their rates (the frequency with which their respective lines trigger), available to data managers at \lhcb.

%% file: text/HLT2.tex
\section{\hlttwo details and results}

\begin{figure}[b!]
    \centering
    \includegraphics[width=0.55\linewidth]{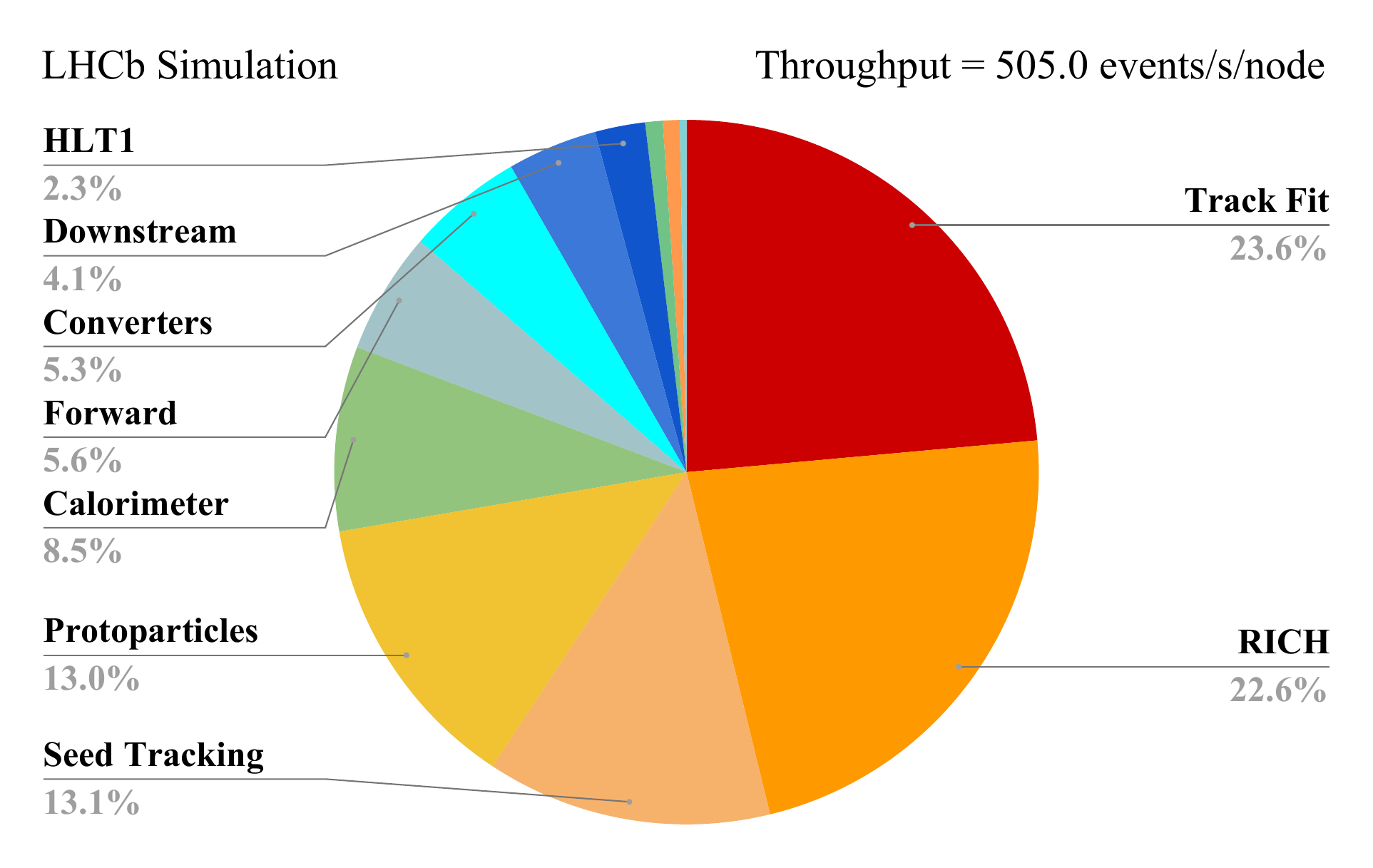}
    \caption{The proportion of time spent by \hlttwo to perform each part of its processes~\cite{LHCb-FIGURE-2022-005}.}
    \label{fig:pie}
\end{figure}

The second level trigger at \lhcb has far lower throughput constraints compared to \hltone and as such may use all the information from the detector to come to its decisions.
This includes using a more accurate tracking algorithm as well as using full PID information from the \rich subdetectors, including further improvements to muon identification.
Each of these processes take 10 times longer than the full \hltone reconstruction (which is also re-run at this stage, using a build transpiled to run on CPUs as a part of \hlttwo).
A pie chart showing how much time is dedicated to each process in \hlttwo is given in \cref{fig:pie}.
The approach to tracking taken in \hlttwo may be specified by the line author to best suit their needs however it defaults to using both forward and seeding-matching methods in full, though this may change in the future.
This differs to the \hltone implementation by using the true magnetic field description rather than the simplified parametric approach.
This is one of the reasons for the track fit being much slower than the one in \hltone instead giving a more-realistic description of the track trajectories.

\begin{figure}[b!]
    \centering
    \includegraphics[width=0.49\linewidth]{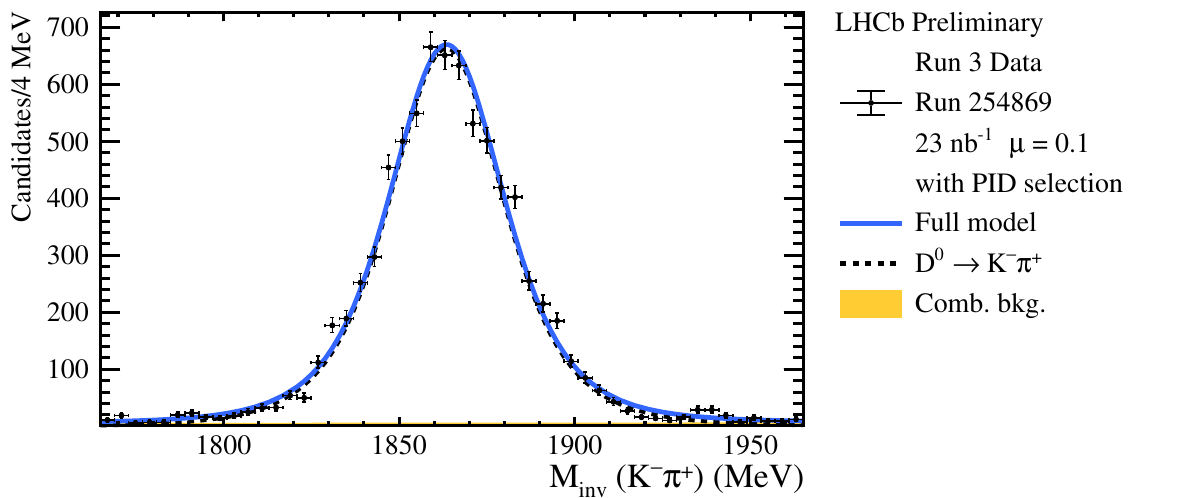}
    \includegraphics[width=0.49\linewidth]{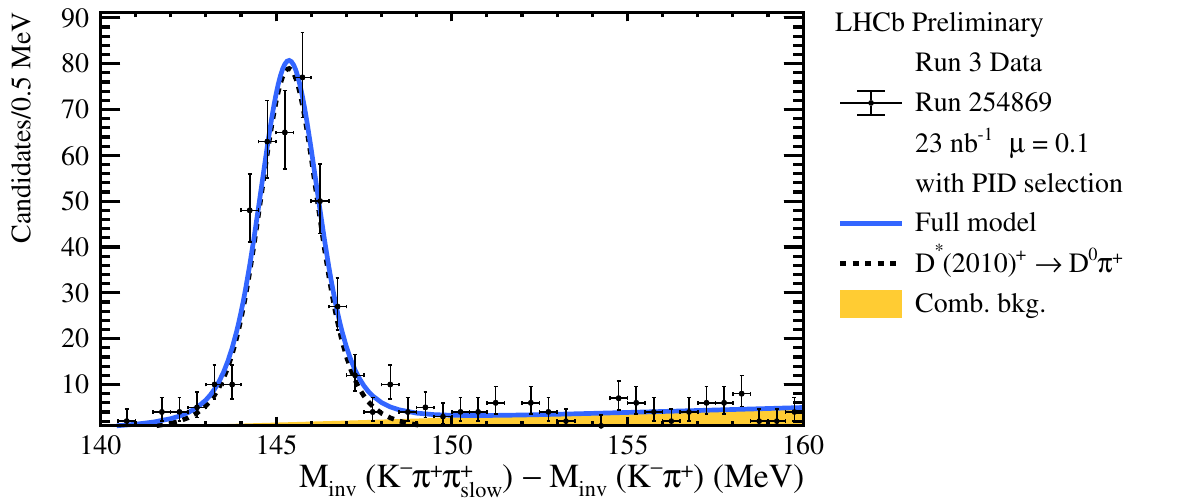}
    \caption{The mass spectra for $\Dz\to\Km\pip$ (left) and $\Dstarp\to(\Dz\to\Km\pip)\pip$ (right) decays using PID information from the \rich subdetectors~\cite{LHCb-FIGURE-2023-002}.}
    \label{fig:D0_hlt2}
\end{figure}

The result of using full PID information as well as better tracking is that far more background may be eliminated from data samples.
This may be seen for $\Dz\to\Km\pip$ decays where the PID allows backgrounds to be reduced to very low levels.
These decays, as well as the excited $\Dstarp\to(\Dz\to\Km\pip)\pip$ decays may be seen in \cref{fig:D0_hlt2}.
Comparing these fits to those from \hltone in \cref{fig:D0} show that dramatic improvements to signal purity may be found.

\begin{figure}[tb!]
    \centering
    \includegraphics[width=0.49\linewidth]{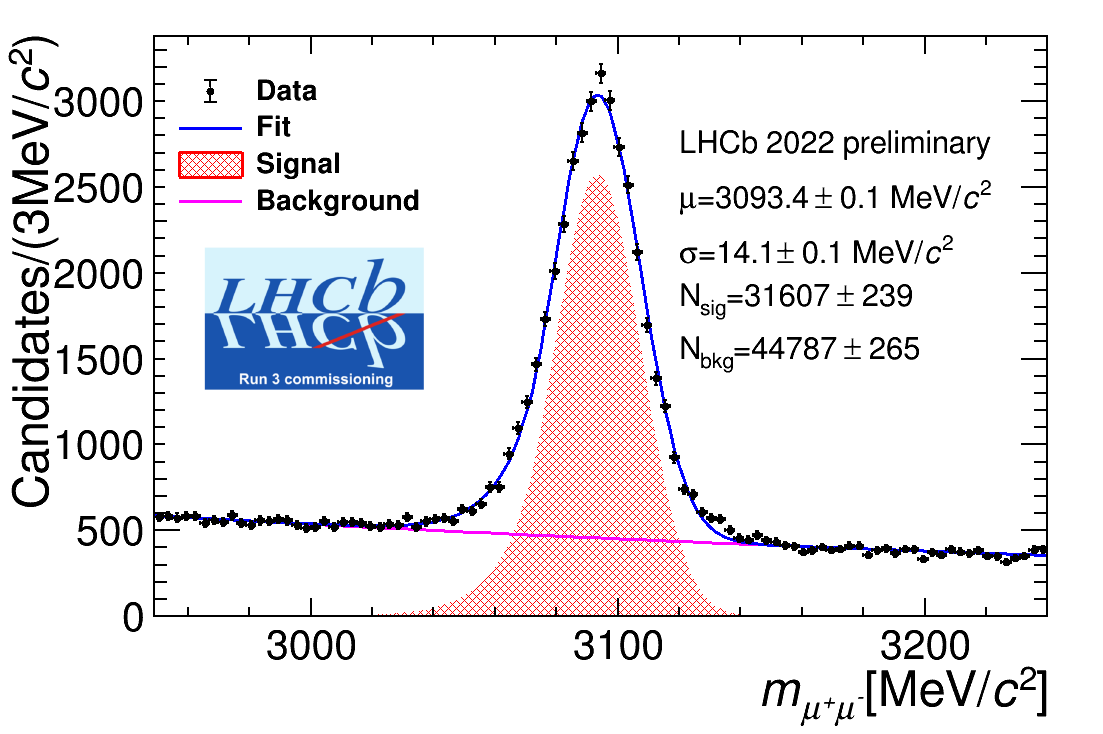}
    \includegraphics[width=0.49\linewidth]{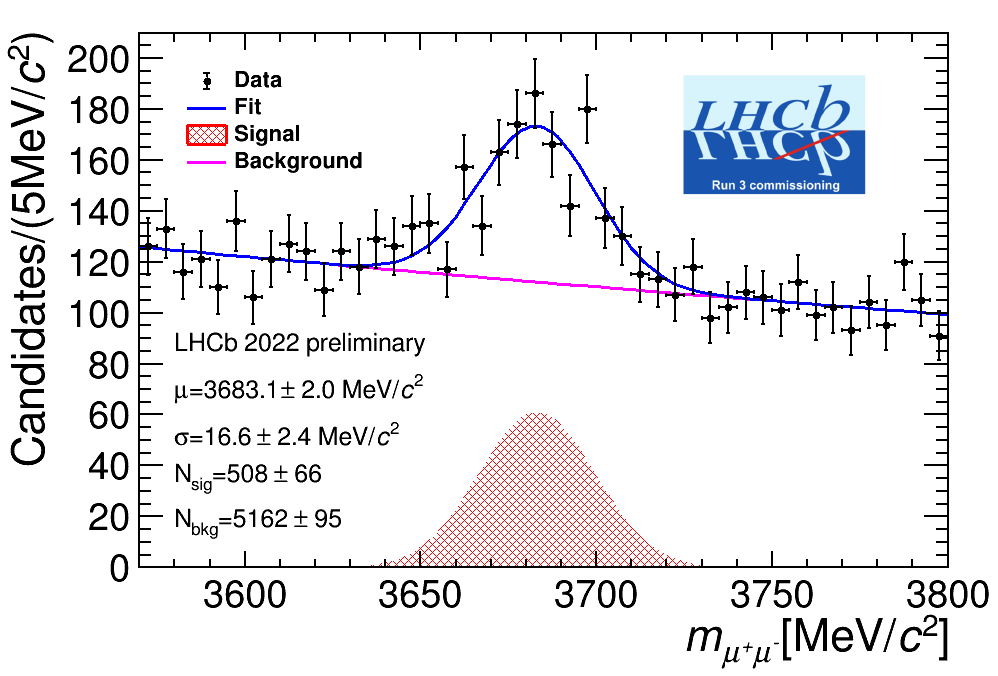}
    \caption{The mass distributions for $\jpsi \to \mup\mun$ (left)~\cite{LHCb-FIGURE-2023-015} and $\psi(2S)\to \mup\mun$ decays~\cite{LHCb-FIGURE-2023-014}.}
    \label{fig:jpsi}
\end{figure}

It is also possible with \hlttwo to find excited charmonium modes with dimuon decays.
Both $\jpsi$ and $\psi(2S)$ decaying to $\mup\mun$ pairs have been found using the dedicated \hlttwo lines.
The results of these studies may be seen in \cref{fig:jpsi}
In this case the background is slightly higher than the hadronic cases shown before, this is due to the muon detector commissioning still being improved at that time.

\begin{figure}[tb!]
    \centering
    \includegraphics[width=0.49\linewidth]{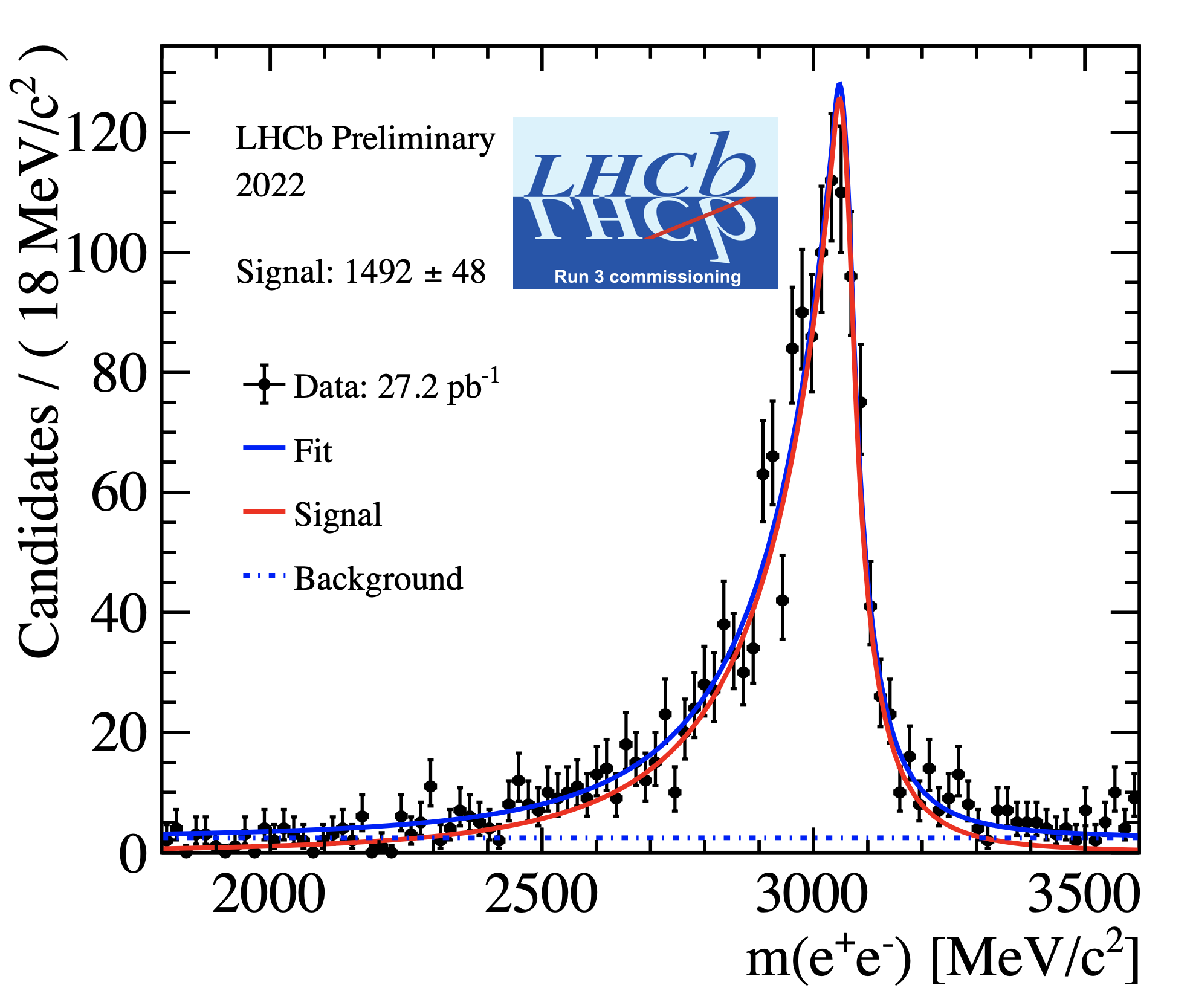}
    \includegraphics[width=0.49\linewidth]{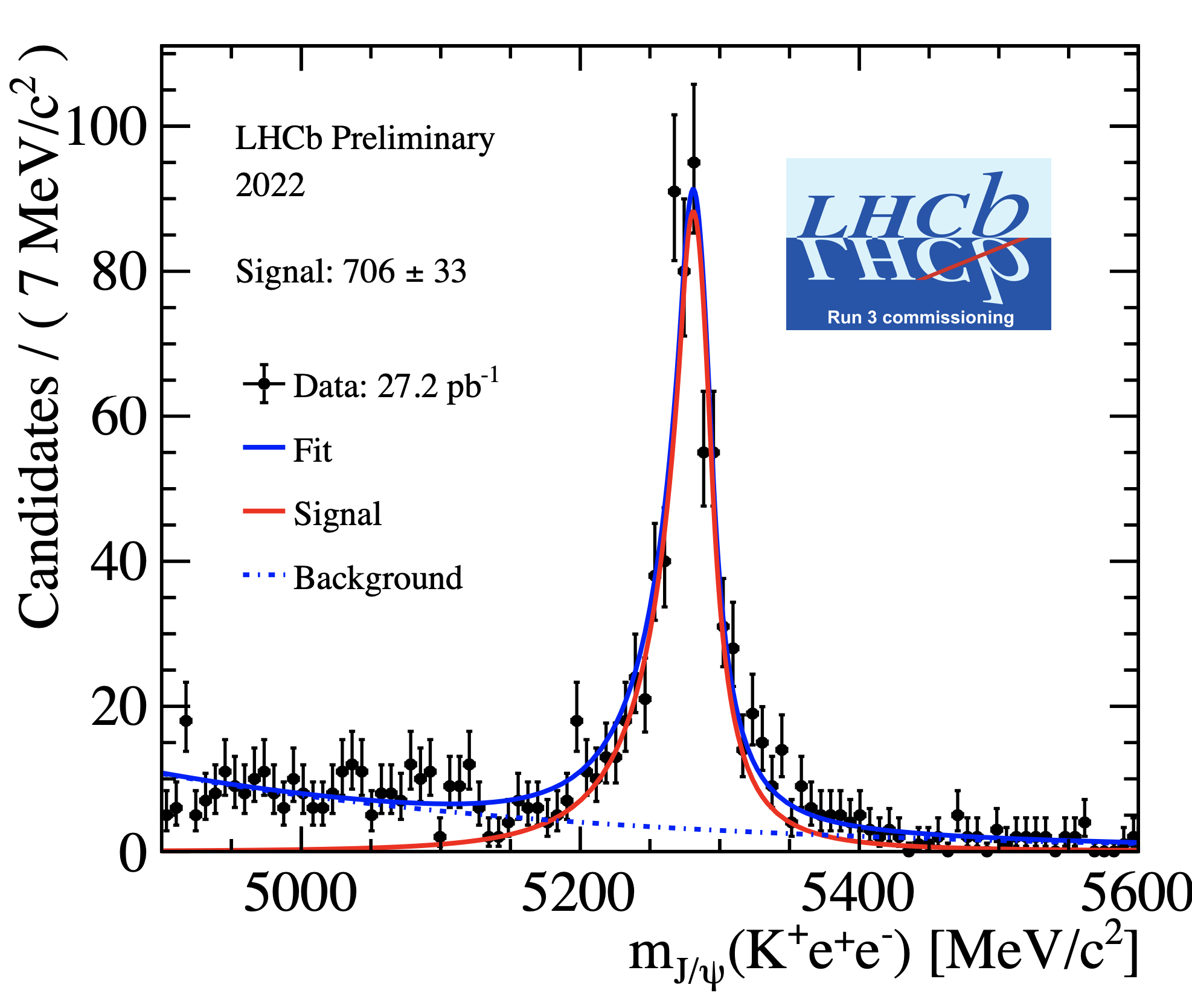}
    \caption{The mass distributions for $\Bp \to(\jpsi \to \ep\en)\Kp$, showing both the \jpsi candidate mass (left) as well as the \Bp candidate mass (right)~\cite{LHCb-FIGURE-2023-010}.}
    \label{fig:BtojpsiK}
\end{figure}

\begin{figure}[tb!]
    \centering
    \includegraphics[width=0.325\linewidth]{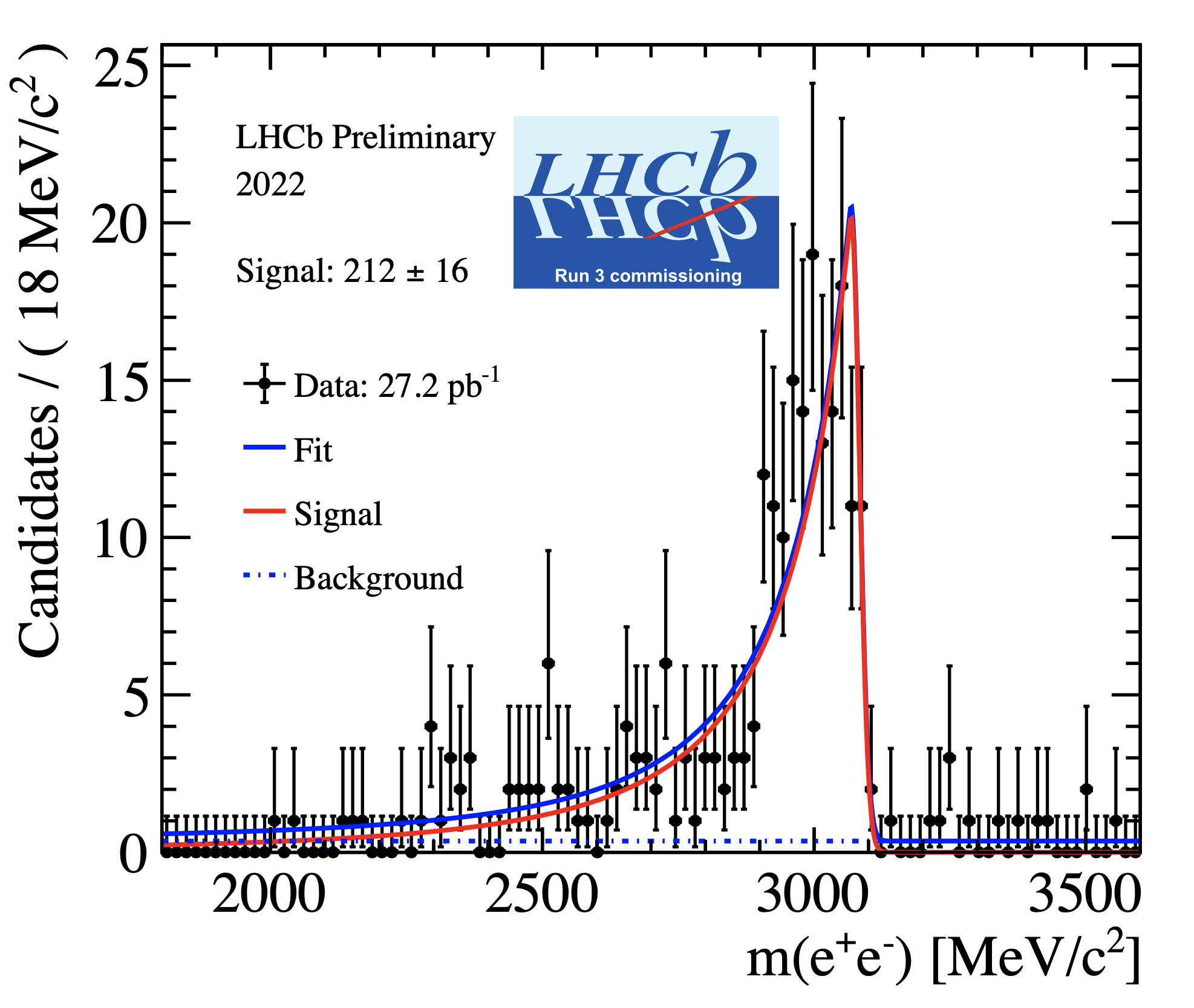}
    \includegraphics[width=0.325\linewidth]{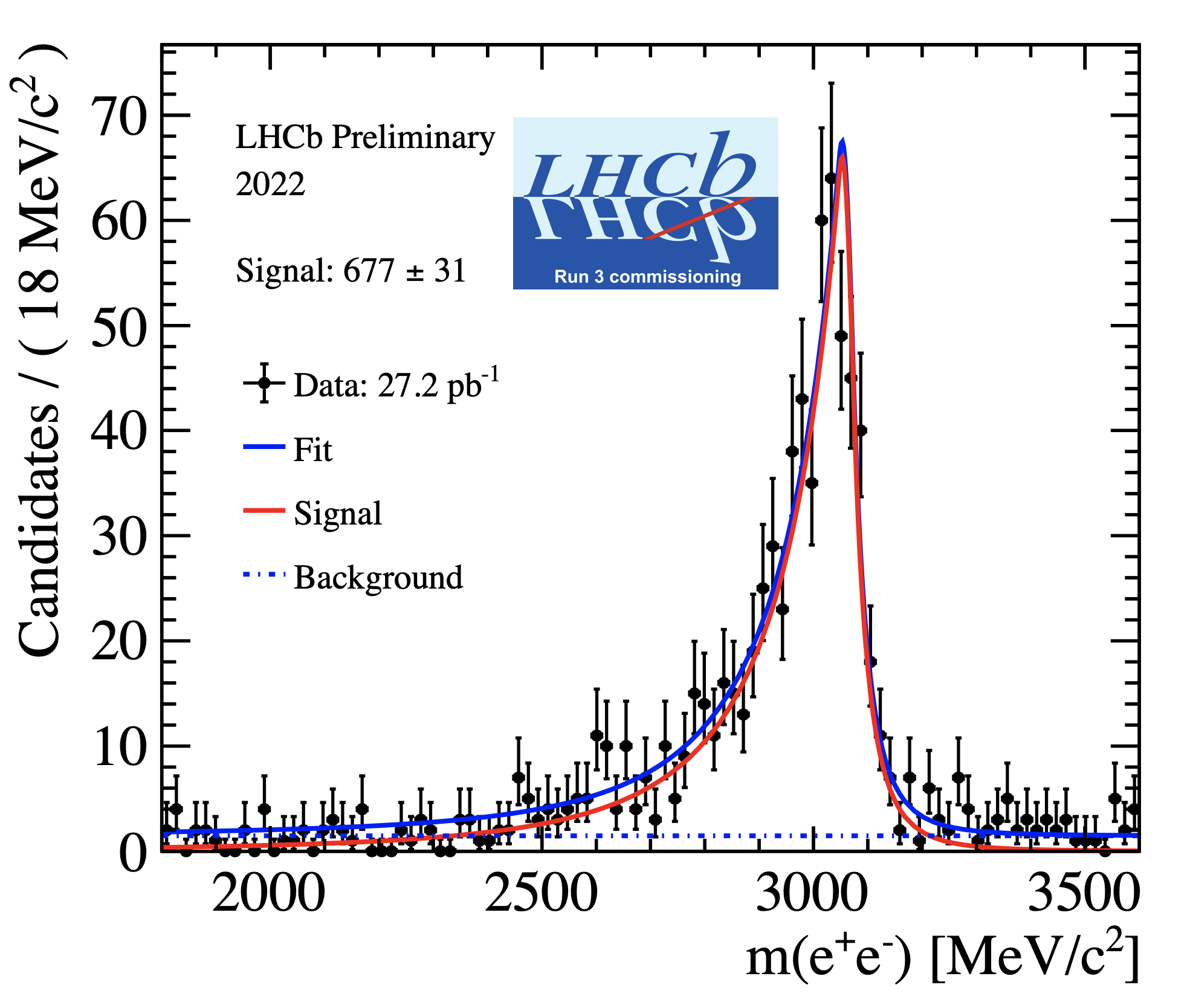}
    \includegraphics[width=0.325\linewidth]{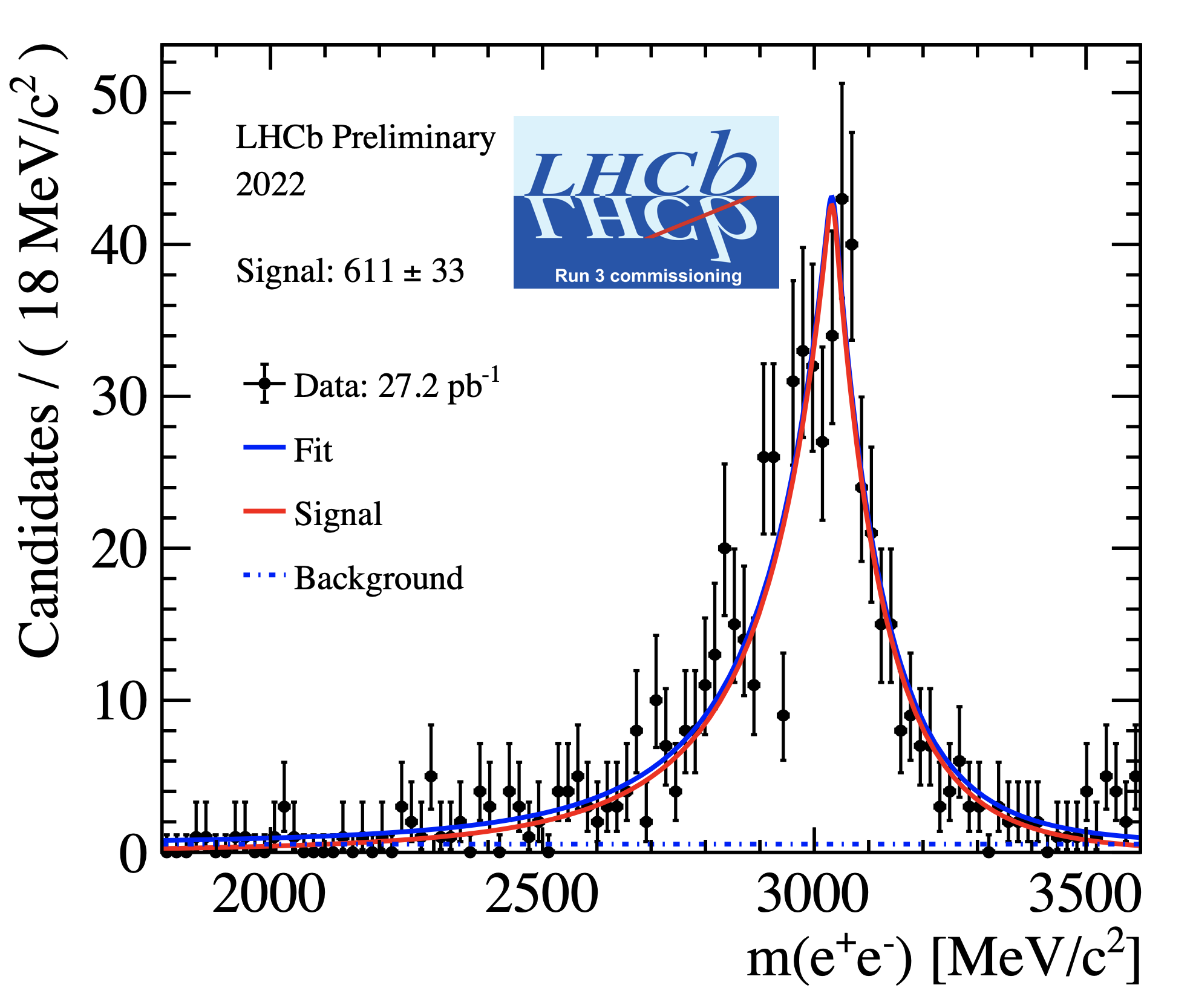}
    \caption{Reconstructed \jpsi masses in categories of bremsstrahlung. Shown here are the distributions for if zero (left), one (center) or more than one (right) photon may be recovered. As more photons can be recovered, the reconstructed energy of the photon better matches the true value and thus the distributions become more symetrical~\cite{LHCb-FIGURE-2023-010}.}
    \label{fig:brem}
\end{figure}

Finally, it has been possible to reconstruct \bquark-hadron decays using \hlttwo, including in electronic modes accounting for effects such as bremsstrahlung.
Decays of $\Bp\to(\jpsi\to\ep\en)\Kp$ have been studied and identified with good purity, shown in \cref{fig:BtojpsiK}.
The reconstructed \jpsi candidates may further be categorised based on if their bremsstrahlung emissions can be recovered.
Bremsstrahlung is an effect where charged particles travelling through either the magnet region or through material of \lhcb will emit soft photons affecting their kinematics, having particularly significant effects on light particles such as electrons. The categories used are for if zero, one, or more than one photon may be recovered by matching clusters made in the \ecal to parent electron tracks. The resulting \jpsi masses in these categories are given in \cref{fig:brem}.

%% file: text/conclusions.tex
\section{Summary \& conclusions}

Overall \lhcb's Real Time Analysis project has, over the course of its commissioning process, provided very good early results though its three-stages.
This is facilitated by firstly the high-throughput heterogeneous \hltone trigger, capable of providing accurate physics results directly from the detector output in real time.
Secondly, these results are then refined by the more-precise \hlttwo trigger, using PID information and better tracking to identify events with higher purity, allowing analysists to perform a majority of their selection within the definitions of their lines, rather than offline.
Finally, the alignment and calibration being provided in real time, improving physics results as data is taken have allowed for optimisation of the detector response.

Using the triggering system it has been possible to reconstruct hadronic signals; leptonic signals; signals with multiple decays; and signals which suffer the effects of bremsstrahlung.
This covers the use cases of almost the entire physics program at \lhcb proving not only that the trigger commissioning is progressing well, but also that it will provide high-quality data to analysists during Run~3.

%% file: references.bbl
\ifx\mcitethebibliography\mciteundefinedmacro
\PackageError{LHCb.bst}{mciteplus.sty has not been loaded}
{This bibstyle requires the use of the mciteplus package.}\fi
\providecommand{\href}[2]{#2}